\documentclass[prl,twocolumn,pslatex]{revtex4}
\usepackage{array,amsmath,amssymb,graphicx}

\begin{document}

\title{Coulomb blockade and transport in a chain of
one-dimensional quantum dots}

\author{Michael M. Fogler}


\affiliation{Department of Physics, University of
California San Diego, La Jolla, 9500 Gilman Drive, California 92093}

\author{Sergey V. Malinin}

\affiliation{Department of Chemistry, Wayne State University, 5101 Cass
Avenue, Detroit, Michigan 48202}

\author{Thomas Nattermann}

\affiliation{Institut f\"ur Theoretische Physik, Universit\"at zu K\"oln,
Z\"ulpicher Str. 77, 50937 K\"oln, Germany}

\date{\today}

\begin{abstract}

A long one-dimensional wire with a finite density of strong random
impurities is modelled as a chain of weakly coupled quantum dots. At low
temperature $T$ and applied voltage $V$ its resistance is limited by
``breaks'': randomly occuring clusters of quantum dots with a special
length distribution pattern that inhibits the transport. Due to the
interplay of interaction and disorder effects the resistance can exhibit
$T$ and $V$ dependences that can be approximated by power laws. The
corresponding two exponents differ greatly from each other and depend
not only on the intrinsic electronic parameters but also on the impurity
distribution statistics.

\end{abstract}

\maketitle


Recently much interest has been attracted by the observations of
algebraic current-voltage and current-temperature dependences in
one-dimensional (1D) electron systems. The power-laws, $I \propto V^{b +
1}$ at high $V$ and $I \propto T^a V$ at low $V$, appear in a variety of
conductors including carbon nanotubes~\cite{Bockrath_99,
Bachtold_01, Gao_04, Tzolov_04}, nanowires~\cite{Zaitsev-Zotov_00,
Chang_03, Tserkovnyak_03, Slot_04, Venkataraman_06}, and polymer
nanofibers~\cite{Aleshin_04}. This implies that the electron transport
in such systems is effectively blocked by some barriers at $T$, $V \to
0$. It is a question of both fundamental and practical interest to
determine what these barriers are and what controls the exponents $a$
and $b$.

In the model with a single opaque barrier the power-law can indeed
appear if electron interactions are strong, in which case the effective
transparency of the barrier is governed by the $T$ and $V$-dependent
suppression of electron tunneling into a correlated many-body
state~\cite{Kane_92}. Similar physics may operate in quasi-1D
case~\cite{Larkin_78, Devoret_00, Ingold_92, Egger_01}. Barring some
difference in computational methods, the existing theories agree that
the single-barrier exponents should be equal. For an interior barrier in
a 1D wire they are given by $a = b = 2 (K^{-1} - 1)$ where $K < 1$ is
determined by the interaction strength~\cite{Kane_92}. Experiments with
wires a few $\mu{\rm m}$ or shorter in length~\cite{Bockrath_99,
Bachtold_01, Gao_04, Tserkovnyak_03, Venkataraman_06} support these
predictions. Yet in wires of length $L \gtrsim 10\,\mu{\rm m}$ the ratio
$a / b$ is almost always larger than unity. It is usually between two
and five~\cite{Zaitsev-Zotov_00, Slot_04, Venkataraman_06, Aleshin_04}. To
explain this behavior we propose that for long wires the
single-barrier model may be unrealistic. Indeed, observations of Coulomb
blockade (CB) at low $T$~\cite{Bockrath_99, Bachtold_01, Aleshin_04}
suggest that a long wire typically contains multiple impurities that
convert it into a chain of weakly coupled 1D quantum dots.

%
%
\begin{figure}
\centerline{
\includegraphics[width=2.85in,bb=190 554 439 755]{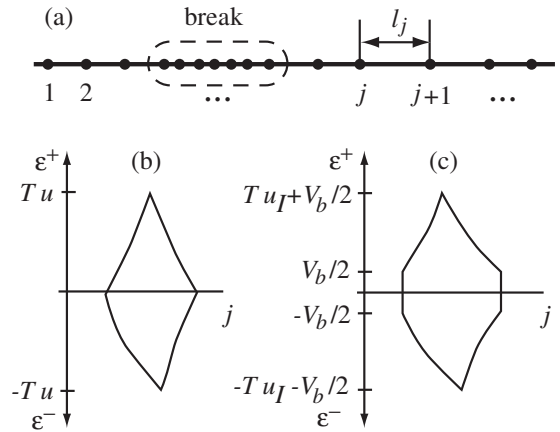}
}
\vspace{0.0in}
\caption{
(a) Illustration of the model. The consecutively numbered dots represent
strong random impurities.
(b) Typical shape of an Ohmic break.
(c) A typical non-Ohmic break.
\label{Fig:breaks}
}
\end{figure}

We show that for such a system there indeed exists a broad parameter
regime where the function $I(V, T)$ can be approximated by power laws
with $a \gg b$. They originate from the statistics of impurity
distribution \emph{not\/} from Luttinger-liquid effects~\cite{Kane_92}.
Although in strictly 1D wires these power laws would be somewhat
obscured by mesoscopic fluctuations, they may be observable in a cleaner
manner in quasi-1D systems~\cite{Zaitsev-Zotov_00, Slot_04,
Venkataraman_06, Aleshin_04}. Under certain realistic conditions
(discussed below) different transverse channels of such systems can act
as independent 1D conductors in parallel, so that our results are fully
relevant yet mesoscopic fluctuations are averaged out.


The model we study is similar to that of Refs.~\cite{Nattermann_03,
Fogler_04, Malinin_04} and is depicted in Fig.~\ref{Fig:breaks}(a).
Identical impurities (e.g., dopants of the same chemical species) are
positioned along the wire according to a Poissonian distribution with
the mean spacing $l \gg 1 / n$, where $n$ is the average electron
density. The tunneling transparency of each impurity is $e^{-s} \ll 1$.
Electrons occupy a partially filled band with a generic (e.g.,
parabolic) dispersion. Their spin is ignored. Electron-electron
interactions are short-range, e.g., due to screening by a nearby gate,
so that the system has a finite (geometry-dependent) capacitance per
unit length $C$. The possibility of thermal activation and a finite
conductance in the CB regime is ensured by weak coupling of electrons
to the thermal bath of phonons. (The conductance depends on this
coupling through a subleading prefactor, if at all~\cite{Gornyi_05}. We
ignore such prefactors here because they are nonuniversal.)

The ground state and the low-energy charge excitations of this model are
well approximated by the classical limit in which the number of
electrons in each dot is restricted to be an integer. In the ground
state it is the integer $q_j$ closest to the ``background charge'' $Q_j
= n l_j + (\xi / \Delta_j)$, where $\xi$ is the electrochemical
potential and $l_j$ is the length of the dot. The charging energy
$\Delta_j = (d \mu / d n + C^{-1}) / l_j$ includes contributions both
due to a finite thermodynamic density of states $d n / d \mu$ (the
analog of the level-spacing in bulk quantum dots) and due to
electrostatics, as usual. Here $\mu = \xi - U$ and $U$ are the chemical
and the electrostatic potentials, respectively, and the units $e = k_B =
1$ are used. The typical charging energy is $\Delta = (d \mu / d n +
C^{-1}) / l$.

Changing the electron number in the dot by $m$ incurs the energy cost of
$E_j^m = (1 / 2) \Delta_j m^2 + \Delta_j m (q_j - Q_j)$. Hence, the
equilibrium probability of having $q_j + m$ electrons in $j$-th dot is
given by $f_j^m = \exp (-\beta E_j^m) / Z$, where $\beta = 1 / T$ and $Z
= \sum_m \exp (-\beta E_j^m)$. The dot is in the CB state if $|q_j -
Q_j| - 1 / 2 \gg T / \Delta_j$. In this case only $m = 0, \pm 1$ are
important. As common in CB literature, one can view $\epsilon_j^\pm
\equiv \mu \pm E_j^{\pm 1}$ as the energy of the lowest unoccupied
(highest occupied) ``single-particle'' orbital. Important in the
following is the probability $P(\epsilon^+, \epsilon^-)$ of having no
charge excitations within a given energy interval, $E_j^{\pm 1} >
\epsilon^\pm$~\cite{Comment_on_spectrum}. Function $P$ depends only on
$\epsilon = \epsilon^+ + \epsilon^-$ and can be easily calculated:
\begin{equation}
 P(\epsilon) = \int \frac{d l_j}{l}
 e^{-l_j / l} \left(1 - \frac{\epsilon}{\Delta_j}\right)
 = 1 - \frac{\epsilon}{\Delta}
 \bigl(1 - e^{-{\Delta}/{\epsilon}}\bigr)
\label{eq:P}
\end{equation}
(the integration is constrained to $\Delta_j > \epsilon$). It behaves as
$\Delta / (2 \epsilon)$ at $\epsilon \gg \Delta$, and so the average
charge excitation gap $\langle \epsilon \rangle = -{\textstyle \int}
\epsilon d P(\epsilon)$ is logarithmically
divergent~\cite{Comment_on_P}. Below we turn to the study of transport
and show that this divergence and the constrains of the 1D geometry lead
to new conductivity laws.

We focus on the case of low voltages, $V < V_c = L \Delta / (2 l)$,
where the wire is in the CB state~\cite{Middleton_93}. Electron
transitions between any two dots $j$ and $k$ occur by the combination of
quantum tunneling and thermal activation. For $|j - k| = 1$ the
tunneling is said to be sequential; otherwise, it is referred to as
cotunneling~\cite{Ingold_92}. In prior studies of random quantum dot
arrays~\cite{Fogler_04, Feigelman_05, Tran_05, Artemenko_05} the analogy
between this transport mechanism and the variable-range
hopping~\cite{Mott_book, ES84} (VRH) has been noted. Thus, the typical
length of a hop $|j - k|$ is established from the competition between
activation and tunneling. Longer hops allow electrons to find
transitions of lower activation energy but involve a higher tunneling
action $s |j - k|$. The crucial point is then as follows. It is known
that 1D VRH (of non-interacting electrons) is special: the conductance
is limited not by typical hops but by rare highly resistive spots ---
``breaks''~\cite{Kurkijarvi_73, Lee_84, Raikh_89, Fogler_05}. Therefore,
one may anticipate that some type of breaks operate in our model as
well. Indeed, the calculation below shows that at not too low $T$ the
role of breaks is played by rare clusters of densely spaced impurities,
see Fig.~\ref{Fig:breaks}(a)~\cite{Comment_on_Larkin}.

The calculation closely parallels those of Refs.~\cite{Raikh_89} and
\cite{Fogler_05}. It starts by noting that if breaks exist, they act as
transport bottlenecks where the most of the applied voltage drops. In
contrast, the dots between two adjacent breaks are in a
quasi-equilibrium. Their electrochemical potentials $\xi_j$ are nearly
equal while the occupation factors of their charge excitations are given
by $f_j^{\pm 1} \simeq \exp[\pm \beta (\mu_j - \epsilon_j^\pm)]$, where
$\mu_j$ is the \emph{local\/} chemical potential. This allows one to get
a simple formula for the net current $I_{j k}$ from a dot $j$ to another
dot $k$. For the most relevant case of a large activation energy, $E_{j
k} \gg T$, it coincides with the standard expression of the VRH theory:
\begin{align}
I_{j k} &\sim I_0 e^{-s |j - k|} e^{-\beta E_{k j}}
\sinh \beta ({\xi_j - \xi_k}),
\label{eq:I_pair}\\
E_{k j} &= \frac12 \min_{\sigma, \tau = \pm}
(|\epsilon_j^\sigma - \mu_j|
 + |\epsilon_k^\tau - \mu_k|
 + |\epsilon_j^\sigma - \epsilon_k^\tau|),
\label{eq:E_pair}
\end{align}
where $I_0 = \text{const}$ is some prefactor~\cite{Comment_on_I}. Consider
first the Ohmic regime. Here $\mu_j \simeq \mu = \text{const}$ and
the effective resistance $R_{j k} = (\xi_{j} - \xi_{k}) / I_{j k}$ for
each link of the hopping network can be defined: $R_{j k} = R_0
\exp(\beta E_{j k} + s |j - k|)$ with $R_0 = T / I_0$.
Denote by $P_u(u)$ the probability per unit length of finding a break of
resistance at least $R_0 e^u$, i.e., such a configuration of dots that
$R_{j k} \geq R_0 e^u$ regardless of how $j$ to the left of the break
and $k$ to the right of it are chosen. For $u \gg s$ this break contains
many, at least $u / s$, dots. We can then talk about its
\emph{shape\/}, i.e., smooth envelope functions $\varepsilon^\pm(z) > 0$
such that $\pm (\epsilon_j^\pm - \mu) \geq \varepsilon^\pm(j)$ for all
$j$ in the break, see Fig.~\ref{Fig:breaks}(b). The probability in
question can be approximated as follows:
\begin{equation}
P_u(u) \sim \max \prod_j P[\varepsilon^+(j) + \varepsilon^-(j)],
\label{eq:P_u_product}
\end{equation}
where the maximum is sought under the constraint $R_{j k} \geq R_0 e^u$
for all $(j, k)$ pairs. Verbatim repetition of the steps taken
in Ref.~\cite{Raikh_89} establishes that the break has a familiar
diamond-like shape~\cite{Raikh_89, Fogler_05}, see
Fig.~\ref{Fig:breaks}(b).

It requires some effort to finish the derivation of $P_u$ for an
arbitrary $u$. However, for $T u \gg \Delta$ the final result can be
obtained if the actual shape of the break is approximated by a rectangle
of the same width and height, i.e., if we consider a break of $N = u /
s$ dots with the charging energy $T u$ each. This gives $P_u(u) \sim
[P(T u)]^N$, and so
\begin{equation}
\ln P_u(u) \simeq -(u / s) \ln (T u / \Delta),\quad T u \gg \Delta.
\label{eq:ln_P_u_III}
\end{equation}
A non-Ohmic break, where $V_b \equiv \xi_j - \xi_k \gg T$ for $j$ and
$k$ on the opposite sides of the break, can be treated similarly. The
necessary steps are spelled out in Ref.~\cite{Fogler_05}.
Namely, consider a state of a given fixed current $I =
I_0 \exp(-u_I)$, so that the free parameter is the voltage $V_b$
generated by a break. Denote by $P_V(V_b)$ the probability per
unit length of finding a non-Ohmic break with the voltage drop of at
least $V_b$. Once again approximating the break
shape by a rectangle $\varepsilon^\pm = T u_I + (V_b / \mathcal{L})$
with $\mathcal{L} \sim 1$~\cite{Comment_on_hexagon} and using
Eq.~(\ref{eq:P_u_product}), we find this time
\begin{equation}
\ln P_V(V_b) \simeq \ln P_u(u_I) -  V_b / (s \mathcal{L} T).
\label{eq:ln_P_V_II}
\end{equation}
This entails that the average voltage drop $V_*$ across a non-Ohmic
break is $V_* = s \mathcal{L} T$. Thus, if $V \gg V_*$, many breaks
must contribute. The derivation of $I(V)$ is especially simple in this
case, so we prefer to finish it first and turn to the Ohmic one later.

The total voltage $V$ is the sum of voltages generated by all non-Ohmic
breaks (contribution of Ohmic ones is subleading). For $V \gg V_*$ (many
breaks) the sum can be replaced by the integral
%
$
              V = L \textstyle \int d V_b P_V(V_b)
$.
%
Using Eqs.~(\ref{eq:ln_P_u_III}) and (\ref{eq:ln_P_V_II}) we get a
transcendental equation for $u_I$
\begin{equation}
u_I =  s\, {\ln (c_1 L T / V l)} /\, {\ln (T u_I / \Delta)},
\quad V_* \ll V \ll V_c,
\label{eq:non-Ohmic_I_V}
\end{equation}
where $c_1$ is a slow dimensionless function of $V$ and $T$. We conclude
that the dependence of $I$ on $V$ is close to the power law with
a weakly $T$ and $V$-dependent exponent
\begin{equation}
b \simeq s \,/ \ln ({T u_I} / {\Delta}),
\quad T \gg \Delta / s.
\label{eq:b}
\end{equation}
This is the first of our two main results. As $V$ decreases to a number
ca. $V_*$ the resistance of a wire becomes dominated by the single
largest break, and so it exhibits strong ensemble fluctuations. To
illustrate how to handle this case, let us consider next the Ohmic
regime where such fluctuations are the strongest.

In this regime the resistance $R$ of the wire is the sum of Ohmic
resistances $R_j = R_0 \exp(u_j)$, where $u_j$'s form a set of $N_b \sim
L / l_b$ independent random numbers with the probability distribution
function (PDF) $l_b P_u(u)$ each [Eq.~(\ref{eq:ln_P_u_III})]. Parameter
$l_b^{-1} \equiv {\textstyle \int} P_u(u) du$ enters the final result
only under the logarithm, so it need not be specified precisely.

Following~\cite{Raikh_89}, our goal is to compute $P_R$, the
PDF of $R$. We begin by noting that the PDF $p$ of $R_j$ behaves as
\begin{equation}
p(R_j) \sim R_j^{-1 - \alpha},\quad
\alpha = s^{-1} \ln \left[({T}/{\Delta}) \ln ({R_j}/{R_0})\right],
\label{eq:p}
\end{equation}
cf.~Eq.~(\ref{eq:ln_P_u_III}). Since $\alpha(R_j)$ is a very slow
function, $p(R_j)$ is basically a power-law. Hence, the random variable
$R > 0$ is a L\'evy random walk. A celebrated theorem of the probability
theory immediately tells us that in the limit of large $N_b$ the PDF of
$R$ approaches the stable L\'evy distribution~\cite{Zolotarev_86} (often
encountered in plasma, astro, and atomic physics, as well as in biology,
economics, and reliability theory). For $\alpha \ll 1$ it becomes
identical to Fr\'echet extreme-value statistics $P_R(R) \propto R^{-1 -
\alpha_*} \exp[-(R_* / R)^{\alpha_*}]$. In turn, the PDF of the
conductance $G = 1 / R$ has the form
\begin{equation}
P_G(G) = \text{const} \times G^{\alpha_* - 1} \exp[-(R_* G)^{\alpha_*}],
\label{eq:P_G}
\end{equation}
so that its first two moments are given by
\begin{equation}
\langle G \rangle R_* = \Gamma(1 + \alpha_*^{-1}),
\quad
\langle G^2 \rangle R_*^2 = \Gamma(1 + 2 \alpha_*^{-1}),
\label{eq:G_mean}
\end{equation}
$\Gamma(z)$ being the Euler gamma-function. 

The equations for $\alpha_*$ and $R_*$ are obtained by the standard
procedure~\cite{Zolotarev_86} (see also Ref.~\cite{Raikh_89}):
\begin{equation}
\alpha_* \ln (R_*/ R_0) = \ln (N_b / \alpha_*),
\quad \alpha_* = \alpha(R_*).
\label{eq:alpha_and_R}
\end{equation}
Combined with Eq.~(\ref{eq:p}), they yield
\begin{equation}
\ln (R_*/ R_0) \simeq 
\frac{s\, \ln (L / \alpha_* l_b)}
     {\ln[(T s / \Delta) \ln(L / \alpha_* l_b)]}.
\label{eq:R_*}
\end{equation}
Not surprisingly, the form of Eq.~(\ref{eq:P_G}) is identical to that found
in Ref.~\cite{Raikh_89} for another model; however, the $T$-dependence
of $R_*$ in that case is different from Eq.~(\ref{eq:R_*}). If only a
limited range of $T$ is accessible, which is often the case in the
experiment, the latter can be approximated by a power law with a weakly
varying exponent. In turn, this entails $\langle G \rangle \propto
T^{a}$, where $a$ is given by
\begin{equation}
a = \frac{d \ln \langle G \rangle}{d \ln T}
  \simeq b\, \frac{\ln (c_2 L / l)}{\ln (T s / \Delta)}
\label{eq:a}
\end{equation}
and $c_2$ is an algebraic function of $T$~\cite{Comment_on_b}. This is our
second main result. As emphasized above, $a \gg b \gg 1$.

Equation~(\ref{eq:a}) applies at $T > \Delta / [s \ln (L / l)]$ and for
wires of length $l \ll L \ll l \exp (e^{s})$. The lower bound guarantees
that the wire contains a large number of impurities. The upper bound,
which is likely to be satisfied in practice, ensures
$\alpha_* \ll 1$, so that Eq.~(\ref{eq:P_G}) holds.

The PDF of Eq.~(\ref{eq:P_G}) describes the statistics of $G$ in an
ensemble of wires at a given $T$. However, fluctuations of comparable
magnitude would appear in the \emph{same\/} wire as $T$
varies~\cite{Lee_84, Raikh_89, Hughes_96}. They would be superimposed on
the backbone dependence $G \propto T^a$. According to
Eq.~(\ref{eq:G_mean}), the amplitude of such fluctuations is large in
the limit we considered, $\alpha_* \ll 1$. Let us however estimate how
small $\alpha_*$ can really be. For the experimentally accessible regime
of $\ln(R_* / R) \lesssim 10$ and $N_b$, $L / l$ of a few hundred,
Eq.~(\ref{eq:alpha_and_R}) gives $\alpha_* \sim 0.5$. In this case, the
fluctuations of $G$ are still significant [cf.~Eq.~(\ref{eq:G_mean})],
and so verification of our predictions in truly 1D systems would require
statistical analysis of the data~\cite{Hughes_96} and/or ensemble
averaging.

At this point we draw attention to the fact that the cited
experiments~\cite{Zaitsev-Zotov_00, Slot_04, Venkataraman_06,
Aleshin_04} were actually performed on quasi-1D systems with a large
number $N$ of transverse channels. Suppose that conductances of
different channels are additive and statistically independent, then for
$\alpha_* = 0.5$ the relative fluctuations of the total $G$ will be
$\sqrt{5 / N}$. This amounts to $10\%$ for a representative
number of $N = 500$, and so the power laws should no longer be obscured
by fluctuations. The above scenario certainly needs an experimental
verification. It will not hold for materials where transverse channels
are coupled. However, there are systems in which it is reasonable, e.g.,
those with a \emph{large\/} concentration of neutral (in general,
short-range) impurities and \emph{strong\/} anisotropy of longitudinal
and transverse transport, see Ref.~\cite{Fogler_04}.

Finally, let us comment on the expected values of $a$, $b$, and $s$. For
a 1D wire the result $s = -\ln t + (K^{-1} - 1) \ln(E_F^2 / \delta_j
\delta_{j - 1})$ can be obtained~\cite{Fogler_04, Artemenko_05,
Malinin_04}. Here $t \ll 1$ is the bare tunneling transparency of the
impurity and $E_F$, $\delta_j$ are, respectively, the high and the low
energy cutoffs for the Luttinger-liquid effects. The latter cutoff
$\delta_j = K \Delta_j \propto 1 / l_j$ is the energy spacing of the
collective neutral excitations of the $j$th dot. In comparison, for an
impurity in an infinite wire, the low-energy cutoff is set by either $V$
or $T$~\cite{Kane_92}. Since $s$ depends on $l_j$'s only
logarithmically, our approximation that $s$ is the same for all
impurities is justified. Note that $\delta_j \sim K T u \propto T \ln(L
/ l)$ for important Ohmic breaks, so that, nominally, $s$ has the same
logarithmic $T$-dependence as in the single-barrier case~\cite{Kane_92}.
Nevertheless, due to the (modestly) large logarithmic factors $\ln(E_F /
\delta_j)$ and $\ln(L / l)$ both predicted exponents $a$ and $b$ would
noticeably exceed the single-impurity one, in this case, $2(K^{-1} -
1)$.

Semiconductor nanowires, polymer nanofibers, and carbon nanotubes appear
promising for a continuing investigation of the physics we discussed.
We call for a systematic study of transport in such wires.


We thank the Hellman Fund, the Sloan Foundation, and
Sonderforschungsbereich 608 for support and also A.~Aleshin, D.~Arovas,
C.~Deroulers, and M.~Raikh for valuable comments.



\begin{thebibliography}{99}
\bibitem{Comment_on_spectrum} Each dot also possesses a discrete
spectrum of neutral excitations with a level spacing $\delta_j <
\Delta_j$. This energy scale would show up in a more refined calculation
of $Z$ but more importantly, it sets the infrared cutoff for
the Luttinger-liquid effects, see below.
\bibitem{Comment_on_P} This property holds irrespectively of the precise
form of the dot length distribution as long as it is slowly varying at small
$l_j$, i.e., if close approaches of adjacent impurities are neither too
rare or too common.
\bibitem{Comment_on_Larkin} A similar idea was proposed in
Ref.~\cite{Larkin_78} for a related problem but the competition
between activation and tunneling has not been examined.
\bibitem{Comment_on_hexagon} A more careful calculation along the lines
of Ref.~\cite{Fogler_05} reveals that the dominat breaks are shaped as
hexagons, cf.~Fig.~\ref{Fig:breaks}(b) here and Fig.~1(b) of
Ref.~\cite{Fogler_05}.
\bibitem{Comment_on_b} Theory of Refs.~\cite{Ingold_92},
\cite{Feigelman_05}, and \cite{Tran_05} suggests that the
pre-exponential factor neglected in Eq.~(\ref{eq:I_pair}) may produce
only subleading corrections to Eqs.~(\ref{eq:b}) and (\ref{eq:a}).
\end{thebibliography}
\end{document}